\documentclass[preprint,aps,showpacs,preprintnumbers]{revtex4}
\usepackage{graphicx}
\newcommand{\unit}[1]{~\mathrm{#1}}

\begin{document}

\title{Scaling Law in Carbon Nanotube Electromechanical Devices}
\author{R. Lef\`{e}vre$^{1}$, M. F. Goffman$^{1*}$, V. Derycke$^{1}$,
C.Miko$^{2}$, L. Forr\'{o}$^{2}$, J. P. Bourgoin$^{1}$ and \ P. Hesto$^{3}$}

\affiliation{$^{1}$\textit{Laboratoire d'Electronique
Mol\'{e}culaire, CEA-DSM SPEC, CEA Saclay, 91191 Gif-sur-Yvette, France}\\
$^{2}$\textit{EPFL, CH-1015, Lausanne, Switzerland}\\
$^{3}$\textit{Universit\'{e} Paris 11, Institut d'Electronique Fondamentale,
CNRS, UMR 8622, F-91405 Orsay, France}}

\begin{abstract}
We report a method for probing electromechanical properties of
multiwalled carbon nanotubes(CNTs). This method is based on AFM
measurements on a doubly clamped suspended CNT electrostatically
deflected by a gate electrode. We measure the maximum deflection as
a function of the applied gate voltage. Data from different CNTs
scale into an universal curve within the experimental accuracy, in
agreement with a continuum model prediction. This method and the
general validity of the scaling law constitute a very useful tool
for designing actuators and in general conducting nanowire-based
NEMS.
\end{abstract}

\altaffiliation[Corresponding author ]{ email address: goffman@cea.fr}

\maketitle

Carbon nanotubes (CNTs) are promising candidates for designing and
developing nanoelectromechanical systems (NEMS) because they combine
excellent electronic and mechanical properties. High conductivity of
CNTs allows for designing simple sensing and actuation systems based
on the direct electrostatic coupling with metallic gates. Their
exceptional stiffness, low mass and dimensions ensure operating
frequencies in the GHz range making them suitable for a number of
applications\cite{cleland:2003}.
 Some prototypes of CNT-based NEMS
such as memory devices\cite{Ruekes:2000},
nanotweezers\cite{Kim:1999}, high frequency
oscillators\cite{Sazonova:2004} and actuators\cite{Baughman:1999}
have already been demonstrated.
Theoretical studies of CNT-based switches have been recently published\cite%
{Dequesnes:2002}\cite{Kinaret:2003}\cite{Sapmaz:2003}. Particularly
interesting are molecular dynamics simulations of Dequesnes and coworkers%
\cite{Dequesnes:2002} which have made evident the validity of
continuum models (beam theory) in describing CNT deformation when
its length/diameter ratio is larger than 10. This is a very
important point because it simplifies the description of CNT
mechanical deflection avoiding expensive and time consuming
atomistic simulations. However, designing CNT-based NEMS requires a
very precise knowledge of the static deflection under actuation.
Since in most of the CNT-based NEMS actuation is electrostatic the
relevant parameters are: the geometry of the CNT (inner and outer
diameter, length), its Young's modulus $Y,$ and the geometry of the
device which conditions the CNT- gate(s) electrostatic coupling and
thus the actuation efficiency. The interplay between the different
geometrical parameters indeed makes theoretical prediction tools
indispensable to properly scale any practical device based on
suspended CNTs. Comparison with experiments is equally important
both to evaluate the validity of the predictions and precisely
measure $Y$, since it determines most of the static and dynamic
properties of CNTs.

In this letter, we report an on-chip test method for measuring the
deflection of suspended and electrostatically actuated CNTs. We
develop a theoretical framework for the modelisation of the
electromechanical behavior of CNTs based on continuum models and
validate it experimentally using that test method. Furthermore, we
determine the CNT-Young's
modulus\cite{Treacy:2003}\cite{Poncharal:1999} \cite{Wong:1997}
\cite{salvetat:1999}\cite{Demcsyk:2002} very precisely. The heart of
the system is a doubly clamped suspended CNT three-terminal device
(see Fig.\thinspace 1) deflected by an electrostatic force induced
by a back gate. In this geometry, Van der Waals
forces\cite{Dequesnes:2002} can be neglected. The CNT deflection
only depends on its physical properties and on the electrostatic
environment given by the connecting and gate electrodes, thus
allowing a quantitative comparison with calculations. We measured
the maximum deflection $u_{MAX}$ ($u_{MAX}\equiv H-\mbox{y}(L/2)$ )
see Fig.\thinspace 1a) of a suspended multiwalled CNT as a function
of the applied gate voltage $V_{G}$. We show that data from
different CNTs (with different diameters $D$ and lengths $L$ ) scale
into an universal curve within experimental uncertainties. This
scaling law that we derived from continuum beam theory when
$u_{MAX}\ll D$, can be extended to the $u_{MAX}\geq D$ range, where
the induced stress $T$ due to elongation of the CNT becomes
important. Furthermore it is also valid in the general case where
there is an electrostatic force profile applied along the CNT. The
set of consistent data allows us to accurately determine  $Y$ =
0.41$\pm $0.05 TPa. Our method and the general validity of the
scaling law constitute a very important outcome for designing
conducting nanowire-based NEMS.

A schematic diagram and a typical SEM picture of a CNT-based device
is presented in Fig.\thinspace 1. A multiwalled CNT of diameter $D$
is clamped by two metallic pads and suspended over a length $L$ on
top of a highly doped silicon substrate that acts as a gate. The
distance between the CNT and the gate is fixed by the sacrificial
silicon dioxide layer thickness $H$ ($230$ nm in the present case).
The multiwalled CNTs used were synthesized by arc-discharge
evaporation and carefully purified to remove amorphous carbon and
graphitic nanoparticles\cite{Bonard:1997}. The sample were
fabricated as follows: some droplets of a sonicated suspension of
CNTs in dichloroethane ($50$ $\mu\unit{g}/\unit{ml}$) are deposited
on an oxidized Si wafer, and blown dry after one minute under
nitrogen flow. Atomic force microscopy\cite{scanning:1999} is used
to image, select, and locate CNTs with respect to pre-patterned
alignment marks. Selected CNTs are then connected with two
electrodes designed by electron beam lithography, and deposited by
thermal evaporation of gold (70 nm) with a chromium adhesion layer
(0.5 nm). These metallic electrodes are used both as conducting
electrodes and anchor pads. The sample is dipped in BHF to remove all the SiO%
$_{2}$ underneath CNTs, rinsed in DI water and ethanol and dried on
a hot plate at $50^{\circ}$C.

The sample is mounted in a commercial AFM equipped with a conducting
tip. CNTs and AFM tip are grounded, in order to avoid any
electrostatic interaction between them\cite{general:1}. To measure
the deflection of the CNT when a voltage $V_{G}$ is applied to the
back-gate electrode, the AFM is used in the tapping mode. The
experimental setup is depicted in the insert of Fig.\thinspace 2.
The AFM tip is placed and immobilized\cite{Scanning:1} at the center
of the CNT. The feedback voltage applied to the piezoelectric element $%
V_{piezo}$, that controls the vertical position of the tip, is monitored as
a function of $V_{G}$. Fig.~2 shows a typical result for a suspended CNT, $%
600$ $\unit{nm}$ long and $10$ $\unit{nm}$ in diameter (device {\#}1). As $%
\left| {V_{G}}\right| $ increases, the CNT deflects downwards, the
AFM tip has more room to oscillate and the oscillation amplitude
increases. To keep constant the oscillation amplitude of the tip, a
positive $V_{piezo}$ is applied by the AFM electronics. The measured
$V_{piezo}$, proportional to the deflection of the CNT, is found to
vary as $V_{G}^{2}.$ This is expected for a CNT that follows the
Hooke's law, because the applied force is proportional to
$V_{G}^{2}$ (see Eq.\ref{eq3} below). The AFM tip position is also
influenced by the electrostatic attraction to the substrate. Indeed,
when the tip is placed on the sacrificial silicon dioxide layer at
the same CNT-gate distance (position 2 in Fig.~2), increasing
$\left| {V_{G}}\right| $ increases the attractive electrostatic
force
between the tip and the substrate, the oscillation amplitude decreases and a negative $%
V_{piezo}$ is applied by the feedback loop. The genuine CNT
deflection is obtained by subtracting from the curve measured at
position 1 the one measured at position 2 previously divided by the
relative permittivity of the silicon dioxide (dotted line in
Fig.2)\cite{tip-gate attraction}.

We repeated the experiment on four devices with different $L$ and $D$
parameters\cite{LD param}. Using the calibration of the piezoelectric element%
\cite{This:1} ($243.8$ $\unit{nm}/\unit{V}$) the deflection of the CNT, $%
u_{MAX},$ can be obtained as a function of the applied $V_{G}$. Fig.~3a
shows results for three different devices: {\#}1 ($L=600$\textit{\ }$\unit{nm%
}$\textit{, }$D=10$\textit{\ }$\unit{nm}$ ), {\#}2 ($L=770$ $\unit{nm}$%
\textit{, }$D=13.5\unit{nm}$ ) and {\#}3 ($L=480\unit{nm}$\textit{, }$D=10$%
\textit{\ }$\unit{nm}$\textit{\ }). The parabolic behavior is made evident
in this Log-Log scale representation (the dotted line is a guide to the eye
that has $V_{G}^{2}$ dependence).

The strong dependence of $u_{MAX}\left( {V_{G}}\right) $ on different CNT
parameters ($L$ and $D$) can be understood using a continuum beam equation%
\cite{Landau:1986} for describing the deflection of the CNT (denoted by $%
\mbox{y(x)})$ when an electrostatic force per unit length $F_{elec}$
is applied. In reduced units ($y=\mbox{y}/D$ and $x=\mbox{x}/L$)
this equation reads

\begin{equation}
\frac{d^{4}y}{dx^{4}}-\frac{TL^{2}}{YI}\frac{d^{2}y}{dx^{2}}=\frac{L^{4}}{YID}F_{elec}\left( {y,D,V_{G}}\right)
   \label{eq1}
\end{equation}

Here $Y$ is the Young's modulus (typically of the order of 1 T$\unit{Pa}$%
\cite{Treacy:2003}-\cite{Demcsyk:2002}) and $I$ the moment of inertia of the
CNT approximated\cite{Iapprox} by $I=\pi D^{4}/64$. $T$ is the stress force
induced by elongation of the CNT when the electrostatic force $%
F_{elec}\left(y,D,V_{G}\right) $ is applied. $T$ can be calculated
by,

\begin{equation}
T=\frac{\pi D^{4}Y}{8L^{2}}\int_{0}^{1}\left( \frac{dy}{dx}\right)
^{2}dx  \label{eq2}
\end{equation}

To solve Eq.\ref{eq1} , we have to assume that the stress force $T$ has a
constant value, which is obtained later from the self-consistent condition (Eq.%
\ref{eq2}). The electrostatic force $F_{elec}\left(y,D,V_{G}\right) $
can be modeled by considering the electrostatic energy $E_{elec}=\frac{1}{2}%
C\left(y,D\right) V_{G}^{2}$ and the principle of virtual work\cite%
{Feynman:1977}, $C\left(y,D\right) $ being the capacitance per unit
length of a CNT over a metallic plane. This force has the form:

\begin{equation}
F_{elec}\left(y,D,V_{G}\right) =\frac{2\pi \varepsilon _{0}V_{G}^{2}%
}{D}g\left(2y\right)   \label{eq3}
\end{equation}

Where $\varepsilon _{0}$ is the permittivity of vacuum and $g\left(
t\right) =\left[ \sqrt{t\left( t+2\right) }\ln ^{2}\left(
1+t+\sqrt{t\left( t+2\right) }\right) \right] ^{-1}$ . This function
can be well approximated in the range of practical
interest\cite{our:1} by $g\left( t\right) =A\,t^{-\beta }$ with
$A=0.270$ and $\beta =1.456$ . Since in our case $H\gg u_{MAX}$,
considering the electrostatic force (at a given $V_{G}$) as a
constant is a good approximation. This is strictly true for an
infinite geometry. Indeed, contacting electrodes modify the
electrostatic force near the CNT-metal interface due to screening
and the AFM tip (in spite of being at the same potential as the CNT)
modifies the electrostatic force near its position. We calculated
the electrostatic force profile that builds up by 3D numerical
simulations (FEMLAB) of our structures. It turns out that the
deflection of the CNT produced by the electrostatic force profile is
equivalent to the one generated by a constant force
$F_{elec}(H,D,V_{G})$ times a correcting factor. This factor
$C_{FP}$, is: 1.26 for device {\#}1, 1.28 for {\#}2, 1.45 for device
{\#}3 and 1.21 for device {\#}4 ( $L$ $=730$ $\unit{nm}$, \textit{\
}$D=15\unit{nm}$) \cite{Details:1}.

It is interesting to note that the second term in Eq.\ref{eq1} is
negligible\cite{Landau2} at low $V_{G}$, where $u_{MAX}\ll D$. In
this limit the maximum deflection is well approximated by

\begin{equation}
\frac{u_{MAX}}{D}=K\left( \frac{V_{G}L^{2}}{D^{3-\beta /2}}\right)
^{2}\text{ \ \ , \ \ }K=\frac{C_{FP}A\varepsilon _{0}\left(
2H\right) ^{-\beta }}{3Y} \label{eq5}
\end{equation}

Fig.~3 depicts the results on the four devices rescaled using
Eq.\ref{eq5}. The curves indeed collapse onto a single curve within
the experimental accuracy in determining $L$ and $D$. This finding
implies that $Y$ is nearly the same for all the devices
investigated, in agreement with theoretical
predictions\cite{Jian:1997}. Furthermore, it allows a quantitative
determination of the Young's modulus as shown in the insert of
Fig.~3b, where the points represents $Y$ calculated from the data
using Eq.\ref{eq5} for the four devices investigated. The average
$Y$ value obtained from these experiments is $0.41\pm 0.05$
T$\unit{Pa}$. This value is smaller than expected compared to
previous measurements, although it is statistically
indistinguishable from the one obtain by J. P. Salvetat \textit{et
al.} \cite{salvetat:1999} on the same source of CNTs. One possible
explanation might be defects induced either by the purification
treatment or by the lithographic steps. Experiments are underway to
examine these issues.

In order to explore the validity of the scaling law predicted and
experimentally observed at low deflections, we solved Eq.\ref{eq1} on a
larger $V_{G}$ range. Fig.~4a shows the calculated maximum deflection of the
CNT ($u_{MAX}$) as a function of $V_{G}$ for the structure depicted in
Fig.~1a. Different curves correspond to different $D$, $L$ parameters
(experimentally accessible) for a fixed $H$ value. Since $H$ is still much
greater than $u_{MAX}$, considering the electrostatic force along the CNT as
a constant remains a good approximation. Notice the large variation of $%
u_{MAX}\left( V_{G}\right) $ when $L$ or $D$ varies. This simply
reflects the strong dependence on geometrical parameters. Now, if
the curves depicted in Fig.~4a are rescaled according to
Eq.\ref{eq5} a collapse onto a single curve is made evident (see
Fig.~4b). This scaling law, expected at low $V_{G} $ as shown above,
proves to be valid in the whole range investigated. This striking
result observed for a constant force per unit length can be
generalized to the case of multiple gates as long as the functional
form of the force profile along the CNT is the same for devices with
different $L$ and $D$\cite{Details:1}.

In conclusion, we have presented an accurate method for probing the
electromechanical properties of suspended nanotubes and measuring
the Young's modulus using a conducting AFM on a simple test bench
structure. In this method the maximum deflection $u_{MAX}$ of the
CNT is measured as a function of an electrostatic force controlled
by $V_{G}$. It was shown that $u_{MAX}\left( V_{G}\right) $ can be
rescaled into an universal curve that only depends on the
geometrical parameters of the structure ($L,D$ and $H$) and on $Y$.
This scaling law, together with our on-chip method constitutes a
very useful tool for designing actuators and in general CNT-based
NEMS where a precise knowledge of static deflection is required. We
note that the principle of the method used here can be extended to
other geometries like that of suspended cantilever or multiple gate
structures.

We thank D. Duli\'{c} for critical reading of the manuscript. The research
has been supported by ACI ''nanosciences et nanotechnologies'' programs
(CNRS) and the Swiss National Science Foundation and its NCCR ``Nanoscale
Science''.

\begin{figure}
  \includegraphics[width=7.8cm]{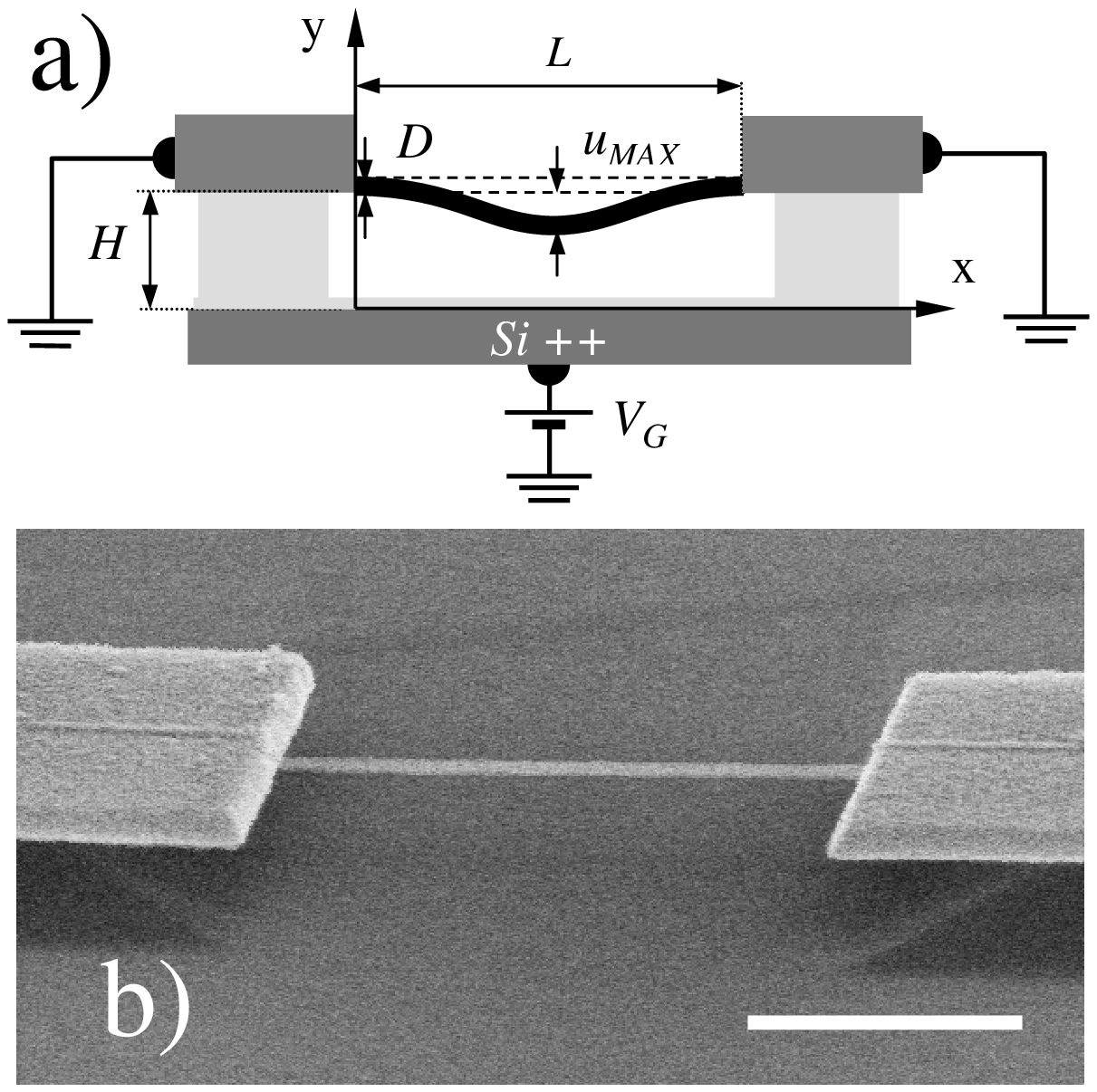}\\
\caption{ a) Schematic picture of a nanoelectromechanical doubly
clamped suspended CNT three terminal device. The distance between
the CNT and the back gate electrode is labeled as $H$. $L$ denotes
suspension length and $D$ the external diameter of the CNT. The
deviation from the straight line is denoted by $\mbox{y(x)}$. b) SEM
image of a typical device used in AFM experiments (scale bar: $500$
$\unit{nm}$). The CNT is connected with two metallic electrodes,
used both as conducting
electrodes and anchor pads. The heavily n doped Si substrate ($10^{19}$ $%
\unit{cm}^{-3}$) act as a back-gate.}
\end{figure}

\begin{figure}
\includegraphics[width=7.8cm]{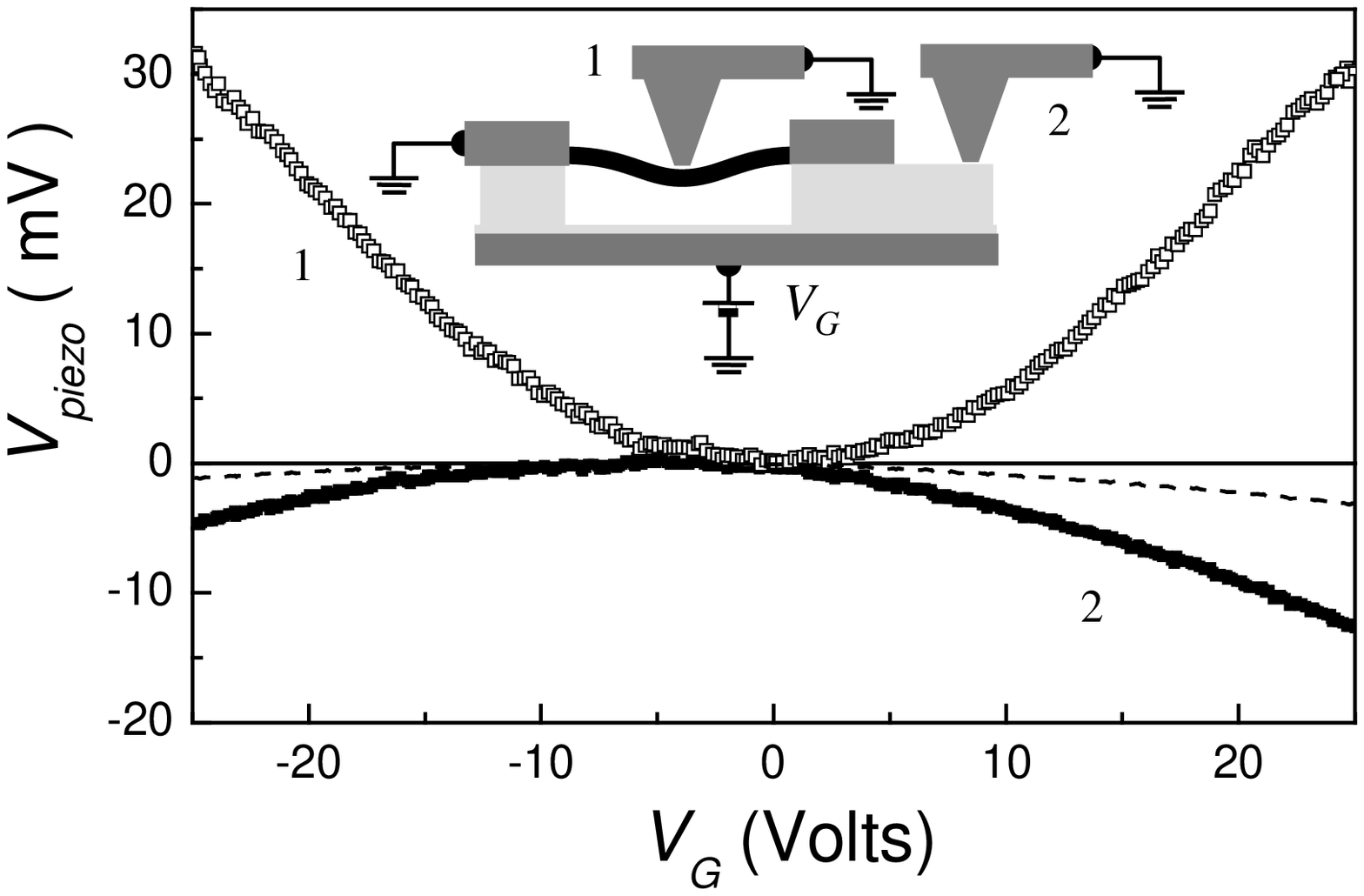}\\
\caption{Feedback voltage applied to
the piezoelectric element $V_{piezo}$ as a function of the gate voltage $%
V_{G}$ at two different positions on device $\#1$ ($L=600$\textit{\ }$\unit{%
nm}$\textit{, }$D=10$\textit{\ }$\unit{nm}$ ): $1$ at the middle of
the CNT ; $2$ on the sacrificial oxide of the silicon substrate (see
insert). In the first case, increasing $\left| V_{G}\right| $
produces the deformation of the CNT, the cantilever tip has more
room to oscillate and the oscillation amplitude increases. To
maintain constant the oscillation amplitude of the tip (tapping
mode), a positive $V_{piezo}$ is applied by the feedback loop.
Conversely, in the second case, increasing $\left| V_{G}\right| $
increases the attractive electrostatic force between the tip and the
substrate, the oscillation amplitude decreases and a negative
$V_{piezo}$ is applied by the feedback loop. The actual CNT
deflection is obtained by subtracting from the curve measured at
position 1 the one measured at position 2 previously divided by the
relative permittivity of the silicon dioxide (dotted line).}
\end{figure}

\begin{figure}
\includegraphics[width=7.8cm]{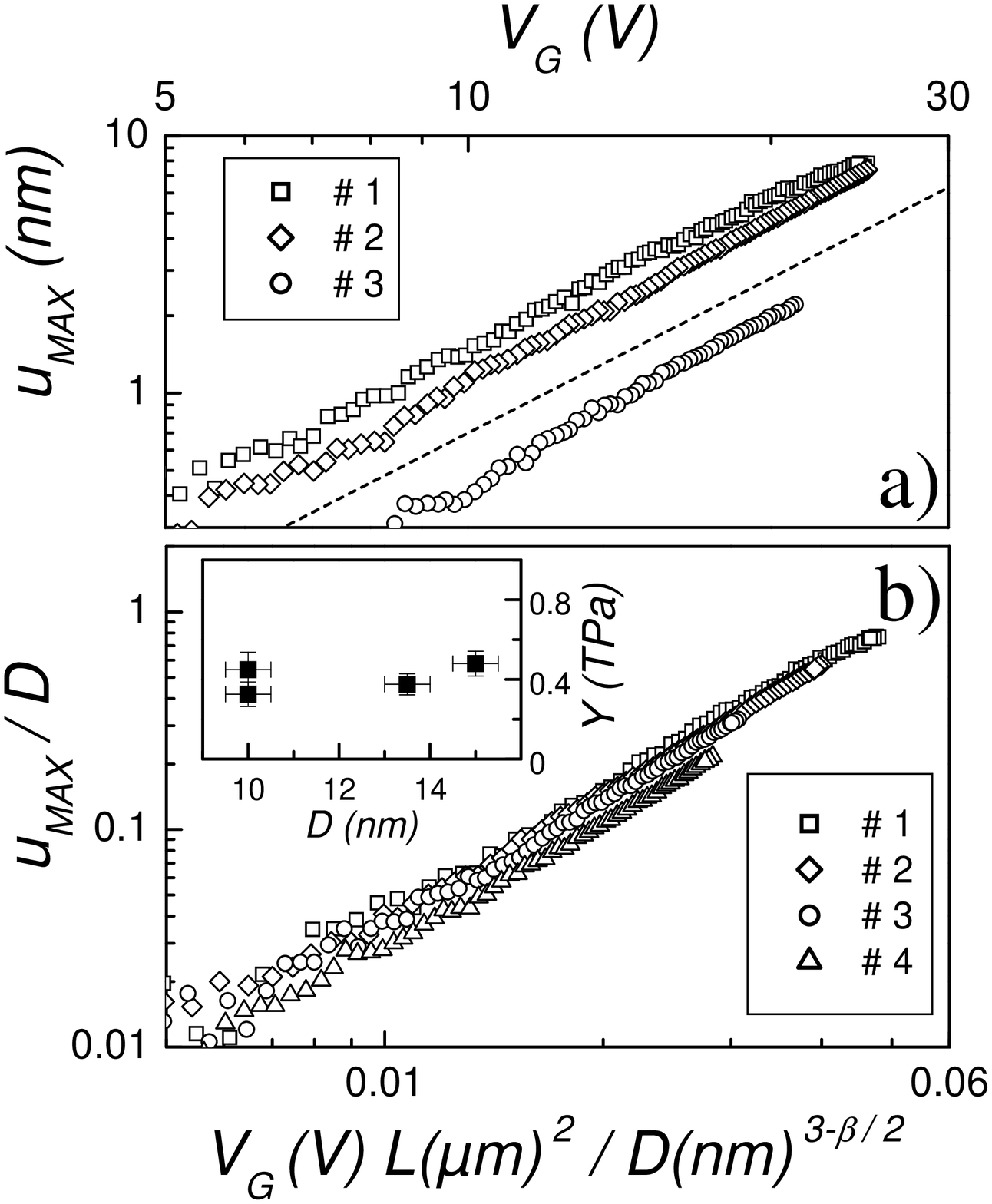}\\
\caption{a) Maximum deflection $u_{MAX}$
as a function of $V_{G}$ for devices $\#1$ ($L=600$ $\unit{nm}$, $D=10$ $%
\unit{nm}$ ), $\#2$ ($L=770$ $\unit{nm}$, $D=13.5\unit{nm}$ ) and $\#3$ ($%
L=480\unit{nm}$, $D=10$ $\unit{nm}$ ). Dash line: guide to the eye that has $%
V_{G}^{2}$ dependence. b) $u_{MAX}$ as a function of $V_{G}$ rescaled using
Eq.\ref{eq5} (see text) for four different devices: $\#1$,$\#2$,$\#3$ and $%
\#4$ ($L=730\unit{nm}$, $D=15$ $\unit{nm}$ ). Insert: Calculated Young's
modulus obtained from the scaling constant $K$ (Eq.\ref{eq5}) for the four
devices. Error bars are estimated from the uncertainty on $L$ and $D$. }
\end{figure}

\begin{figure}[h]
\includegraphics[width=7.8cm]{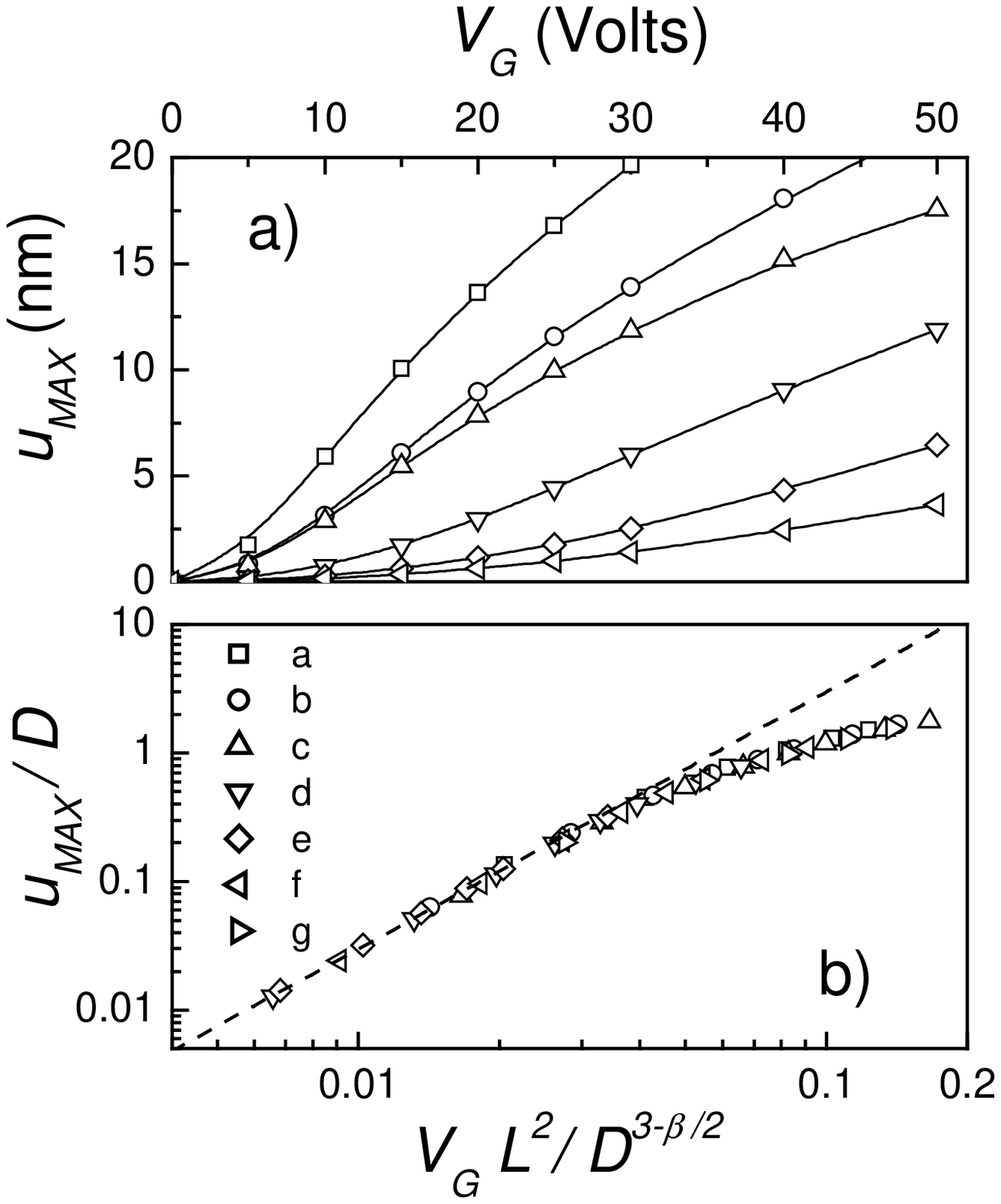}\\
\caption{a) Calculated maximum deflection $u_{MAX}$ as a function of
$V_{G}$ for different $L$, $D$ parameters: a ($L=1200$ $\unit{nm}$, $D=13\unit{nm}$), b (%
$L=1000$ $\unit{nm}$, $D=13\unit{nm}$), c ($L=800$ $\unit{nm}$, $D=10\unit{nm%
}$), d ($L=800$ $\unit{nm}$, $D=15\unit{nm}$), e ($L=800$ $\unit{nm}$, $D=20%
\unit{nm}$), f ($L=500$ $\unit{nm}$, $D=13\unit{nm}$), g ($L=800$ $\unit{nm}$%
, $D=8\unit{nm}$). b) Same results rescaled according to
Eq.\ref{eq5} (see text). Dash line: Analytical result obtained for
induced stress $T=0$.}
\end{figure}


\begin{thebibliography}{99}
\bibitem{general:1} In general a static charge can be present even when the
tip and the CNT are at the same ground potential. This charges, related to
the difference between work functions of the materials involved, are
equilibrated each time the tip and the CNT enter into contact during the
oscillation of the tip.
\bibitem{Scanning:1} Scanning in the plane is switched off. The system is stable enough to ensure a reproducible result.
\bibitem{tip-gate attraction} We proved this fact by
performing an experiment on the same CNT with two different
cantilevers with different sensitivity to the tip-gate attraction.
\bibitem{LD param} $L$ was estimated from SEM images of the devices and $D$
from AFM images before CNT suspension.
\bibitem{Iapprox} In general $I$ is estimated with $\pi \left(
D^{4}-D_{i}^{4}\right) /64$, where $D_{i}$ is the inner diameter of the CNT.
However the value of $K$ (see text) is insensitive to $D_{i}$ , because when $%
D_{i}/D=1/3$, which is an extreme case, $K$ differs only by $1{\%}$
as compared with $D_{i}=0$\textit{.}
\bibitem{our:1} In our case this region was taken in the $10<y<50$ range.
The slight dependence of $\beta $ on $y$ does not change any conclusion of
our work.
\end{thebibliography}
\end{document}